\documentclass[12pt]{article}
\usepackage{graphicx}% Include figure files
\usepackage{bm}% bold math
\usepackage{epsfig}

\newcommand{\beq}{\begin{equation}}
\newcommand{\eeq}{\end{equation}}
\newcommand{\beqa}{\begin{eqnarray}}
\newcommand{\eeqa}{\end{eqnarray}}

\begin{document}
\def\ii{\'\i}

\title{Observable consequences of pseudo-complex
General Relativity}

\author{P. O. Hess$^{1,2}$ \\
{\small\it
$^1$ Instituto de Ciencias Nucleares, Universidad Nacional 
Aut\'onoma de M\'exico,}\\
{\small\it
Circuito Exterior, C.U., 
A.P. 70-543, 04510 M\'exico D.F., Mexico}\\
{\small\it
$^2$ Frankfurt Institute for Advanced Studies, Johann Wolfgang 
Goethe Universit\"at,}\\
{\small\it
Ruth-Moufang-Str. 1, 60438 Frankfurt am Main, Germany}
}

\maketitle

\abstract{
A review of the {\it pseudo-complex General Relativity} (pc-GR)
is presented, with the emphasis on observational consequences.
First it is argued why to use an algebraic extension and
why the pseudo-complex is a viable one.
Afterward, the pc-GR is
formulated. Posterior, several observational consequences are
discussed, as the perihelion shift of Mercury, 
Quasi Periodic Objects, the emission profile
of accretion discs, the pc-Robertson-Walker model of the
universe, neutron stars and gravitational ring-down modes
of a black hole.
}

%\PACS{}
%\keywords{General Relativity, pseudo-complex extension}

\section{Introduction}
\label{intro}

The {\it General Relativity} (GR) is up to now the
most successful theory of Gravitation. Many observations
have been confirmed, especially in week gravitational
fields as in the solar system \cite{will}. 
The existence of gravitational waves were also confirmed
\cite{abbot1,abbot2} and the shadow of a black holes was
observed in \cite{EHT1,EHT2,EHT3,EHT4,EHT5,EHT6}.
Though, the last two observations are related to 
very strong gravitational fields, we will show that
the pc-GR is also able to describe them. The reason lies in 
the method to deduce masses and distances from the 
observations, which is based on the assumption that 
only GR  describes the observations.

Attempts to extend GR have a long history 
\cite{einstein1,einstein2,born1,born2,caianiello1,caianiello2}, all with different motivations. 
For example, A. Einstein wanted to unify gravity with
electrodynamics \cite{einstein1,einstein2}. On the
other hand, M. Born approached a conceptional problem,
namely that in
Quantum Mechanics coordinates and momenta are treated on
an equal footing, while in GR coordinates are predominant. 
The main motivation for pc-GR
\cite{hess2009,book} was to extend GR via algebraic means and
to search for the possibility to eliminate the
event-horizon. Why this is important? This depends on the
view of the beholder: While some argue that the event horizon 
is just another coordinate singularity, others complain that
from a black hole, even nearby, no information can get out.
The no-hair theorem implies that all information is lost,
violating fundamental concepts. Also the singularity in
its center is bothersome and theories have been proposed 
\cite{string,spin} to remedy it.

Though, the event horizon is only a coordinate
singularity, its origin is the strong gravitational field
and there is no reason to believe that GR is still valid.
Thus, a search for an extension 
of GR is justified when the limits
of the theory are reached.

In this contribution, we will elaborate on the motivations 
for proposing the pc-GR and explain its mathematical structure. Apart from presenting a review on the pc-GR, 
the main body of the text concentrates on observational
consequences.
Some of the examples
were already discussed in earlier publications, but
some are new, as will be noted in its course. We will see,
that several predictions cannot be distinguished from GR,
due to either low resolution, or too small corrections, or
alternative explanations within GR.
In these cases, one has to wait for further, improved
observations.

This contribution is organized as follows: In 
Section \ref{alg} algebraic extensions in general are 
briefly discussed. In section \ref{pc-GR} 
the pc-GR is resumed.
In section \ref{obs} the observational predictions and
consequences are discussed, which includes:
The {\it Perihelion shift of Mercury}, 
{\it Quasi periodic objects}, light emission structure of
{\it accretion discs}, the 
{\it pc-Robertson-Walker cosmology},
{\it neutron stars} and {\it gravitational
ring-down modes} of a black hole. 
Some of the results were obtained earlier and some
are new, as resolving the structure of the pseudo-imaginary
part of the coordinates and the perihelion shift of Mercury.
In \ref{concl}
conclusions will be drawn.

The convention $G=c=1$ is used throughout this contribution,
as the metric signature ($-~+~+~+$)

\section{Algebraic extensions}
\label{alg}

A possible option to modify GR is to extend the 
space-time coordinates $x^\mu$ ($\mu = 0,1,2,3$)
to a different kind, which is 
called an {\it algebraic extension}.
The questions to address are: How many algebraic extensions exist? Which of those are consistent with basic principles?
If an algebraic extension is consistent was
addressed in \cite{kelly} where all kinds of coordinates 
where investigated. A more detailed explanation can be
found in \cite{Adv-HEP,PPNP}.

\begin{table}[h!]
\centering
\caption{
Algebras of various algebraic coordinate 
extensions.
} 
\vspace{0.2cm}
\begin{tabular}{|c|c|c|}
\hline\hline
Algebra & Generators $a_i$ & algebra \\
\hline
Real & $1$ & - \\
complex & $1$, $i$ & $i^2=-1$ \\
pseudo-complex & $1$, $I$ & $I^2=1$  \\
quaternion & $1$, $a_1$, $a_2$, $a_3$ & $a_i a_j = 
-\delta_{ij} + \varepsilon_{ijk} a_k$ \\
hyper-quaternions & $1$, $a_1$, $a_2$, $a_3$ & 
$a_1^2 = a_2^2 =1$,     \\
& &  $a_1 a_2 = -a_3$ \\
\hline
 \end{tabular}
\label{tab1}
\end{table}

An algebraic extension is defined through the mapping of
real coordinates to

\beqa
x^\mu & \rightarrow & X^\mu ~=~ x^\mu + a_i y^\mu
\label{alg1}
\eeqa 
where a sum over $i$ is implicit and the $a^i$ satisfy
the algebra

\beqa
a_i a_j & = & C_{ijk} a_k
~~~,
\label{alg2}
\eeqa
with a sum of repeated indexes.
One example is the complex extension which 
contains, besides
$a_0=1$, $a_1=i$ with $i^2=-1$. Another one is the pseudo-complex
(pc) extension, with the generators $a_0=1$ and $a_1=I$ with
$I^2=1$. The possible algebras are listed in Table 
\ref{tab1}. 

Several modifications of GR,
mentioned in the introduction, 
are related to algebraic extensions,
though not obvious at the first sight. In \cite{PPNP}
these attempts were resumed, as the one proposed in
\cite{caianiello1,caianiello2} which implies a maximal
acceleration, or the more obvious one in 
\cite{mantz}, were the complex extension is proposed.

Most of the algebraic extensions run in a serious problems, when the square of at least one of the $a_i$ is -1. It is shown in
\cite{kelly,PPNP} that in the limit of nearly
flat space the propagator of a gravitational wave
has the wrong sign and corresponds 
to a ghost solution, excluded on physical grounds. 
Therefore,
the main observation of \cite{kelly} is that
apart from the real coordinates, corresponding to GR,
the only ones which do not imply ghost and/or Tachyon 
solutions are the pseudo-complex coordinates.
This is a very important finding because it implies that
the path through algebraic extensions is only possible
through the use of pc-coordinates. 

This is the reason why we stick to this extension and in what
follows we will expose some consequences, theoretical and
observational ones.

\section{Pseudo-complex General Relativity}	
\label{pc-GR}

The pc-GR was introduced in \cite{hess2009}. Since then, several reviews were published in \cite{Adv-HEP,PPNP}.
In \cite{book} the theory, their mathematical structure
and some observational facts can be retrieved and
in \cite{hess2013,hess2014} some observational predictions
were published. One centerpiece in pc-GR is that
around any mass vacuum fluctuations accumulate, whose presence
is a consequence of semi-classical Quantum Mechanical 
calculations \cite{birrell,visser1}. 
These vacuum fluctuations have the nature of a dark energy
and semi-classical Quantum Mechanics calculations
result in a fall-off of the dark energy density
as a function in distance to the mass. Up to now, in  pc-GR a {\it phenomenological ansatz} 
is used, assuming 
a fall-off of the dark energy density as 
$\sim B_n/r^{n+2}$, where $B_n$
describes the coupling of the central mass to the dark energy
and $r$ is the radial distance. The $n$ is a parameter, which was assumed to be 3 in \cite{hess2013,hess2014}, because
it is the next leading order correction to the metric which
still does not contradict solar system observations
\cite{will}. However, in \cite{nielsen1,nielsen2} it 
is shown that n has to be $>3$, 
using the first observed gravitational wave event 
\cite{abbot1,abbot2}. For that reason, we 
assume $n=4$ in what follows, keeping
in mind that this is a {\it phenomenological assumption}.

One important consequence of the accumulation of dark energy 
near a central mass is its recoupling to 
the metric, which
is changed such that no event horizon appears. It 
can rephrased into the principle {\it that
a mass not only curves the space nearby but also changes 
the vacuum properties}, which may give clues on the
quantization of gravity.

The pc-Gr is formulated in analogy of the GR, using instead
of real coordinates $x^\mu$ new ones, namely

\begin{eqnarray}
X^\mu & = & x^\mu + I y^\mu ~,~ I^2 ~=~ 1
\nonumber \\
X^\mu & = & X_+^\mu \sigma_+ + X_-^\mu \sigma_-
\nonumber \\
\sigma_\pm & = & \frac{1}{2}\left( 1 + I \right)
~,~ \sigma_\pm^2 ~=~ 1 ~,~ \sigma_+ \sigma_- = 0
~~~,
\label{pc1}
\end{eqnarray}
where two representations are shown, the one
in terms of $1$ and $I$ and the other one in terms of
the so-called {\it zero-divisor basis}, given by
$\sigma_{\pm}$. The $\sigma_{\pm}$ behave as projectors,
with $(\sigma_\pm )^2=1$ and $\sigma_-\sigma_+=0$.
The last property is important because it allows to
perform calculations independently in each zero-divisor
component \cite{book}.  

The metric has also two components and is of the form

\begin{eqnarray}
g_{\mu\nu}(X) & = & g^+_{\mu\nu}(X_+) \sigma_+ + 
g^-_{\mu\nu}(X_-) \sigma_-
~~~.
\label{pc2}
\end{eqnarray}

The length element is defined in terms of $X^\mu$
in the same manner as it is defined in terms of $x^\mu$ in GR:

\begin{eqnarray}
d\omega^2 & = & g_{\mu\nu} dX^\mu dX^\nu
\nonumber \\
& = & 
\left\{g^S_{\mu\nu}\left[ dx^\mu dx^\nu +
dy^\mu dy^\nu \right]
+ g^A_{\mu\nu}\left[ dx^\mu dy^\nu + dy^\mu dx^\nu\right]\right\}
\nonumber \\
&& 
+ I \left\{ g^A_{\mu\nu} \left[ dx^\mu dx^\nu + dy^\mu dy^\nu
\right] +
\right.
\nonumber \\
&& \left.
g^S_{\mu\nu} \left[ dx^\mu dy^\nu + dy^\mu dx^\nu \right]
\right\}
~~~,
\label{pc3}
\end{eqnarray}
with

\begin{eqnarray}
g^S_{\mu\nu} & = & \frac{1}{2}\left(
g^+_{\mu\nu} + g^-_{\mu\nu} \right)
~,~ 
g^A_{\mu\nu} ~ = ~ \frac{1}{2}\left(
g^+_{\mu\nu} - g^-_{\mu\nu} \right)
~~~.
\label{pc4}
\end{eqnarray}
We have used the expression of $X^\mu$ in terms of
$x^\mu$ and $y^\mu$ for rewriting the length element
$d\omega^2$ in (\ref{pc3}).

Requiring that particles only move along real 
distances,
a constraint of a {\it real} length element is 
imposed,
demanding the factor of $I$ in (\ref{pc3}) to vanish, i.e.,

\begin{eqnarray}
&
\left( \sigma_+ - \sigma_- \right)
\left\{ 
g^A_{\mu\nu} \left[ dx^\mu dx^\nu + dy^\mu dy^\nu
\right] +
\right.
&
\nonumber \\
& 
\left.
g^S_{\mu\nu} \left[ dx^\mu dy^\nu + dy^\mu dx^\nu \right]
\right\} ~ = ~ 0
~~~.
&
\label{pc5}
\end{eqnarray}

It is illustrative to consider the GR-limit, i.e.,
$g^A_{\mu\nu}=0$ and with $g^S_{\mu\nu}=g_{\mu\nu}$
now real, the constriction (\ref{pc5}) reduces to

\beqa
g_{\mu\nu}dx^\mu dy^\nu & = & 0
~~~.
\label{pc6}
\eeqa
This is nothing but the dispersion relation, whose 
solution is $y^\nu \sim u^\nu$, the four velocity. For 
dimensional reasons $y^\nu = l u^\nu$,
with $l$ a length parameter, related to the {\it minimal}
length of the theory.

Here, it is important to note that pc-GR contains a {\it minimal
length}, which is a mere real number and is not affected by
a Lorentz transformation. This simplifies enormously
the mathematical structure of the theory, 
because it does not require
a deformation of the Lorentz transformation, needed in theories with a {\it physical} minimal length. 
{\it This
observation hints to the alternative to formulate
theories which contain a minimal length and still
use Lorentz symmetry}.

In general, the solution of (\ref{pc5}) is quite 
complicated and we will discuss two of those in
an approximate manner: For the pc-Schwarzschild case
and for the pc-Robertson-Walker metric.

The action within pc-GR is defined in analogy to GR, namely

\begin{eqnarray}
S=\int dX^4 \sqrt{-g}\left({\cal R}+2\alpha \right)
~~~,
\label{pcGR1}
\end{eqnarray}
where ${\cal R}$ is the pc-Riemann scalar, defined in the
same way as in GR, and $\alpha$ is a pc-constant for
the pc-Robertson-Walker model and may depend on
the radial distance for the Schwarzschild case. Note,
that also the volume element is pseudo-complex.

Varying this action in each zero-divisor sector, using the
constraint (\ref{pc5}), leads to the equations of motion

\begin{eqnarray}
{\cal R}^\pm_{\mu\nu} - \frac{1}{2}g^\pm_{\mu\nu}{\cal R}_\pm
& = &
8\pi T_{\pm~\mu\nu}^\Lambda
\label{pcGR2}
\end{eqnarray}
in each zero-divisor basis. The $\Lambda$ refers
intentionally to its property as a dark energy.

The $T_{\pm~\mu\nu}^\Lambda$ is given by

\begin{eqnarray}
8\pi T_{\pm~\mu\nu}^\Lambda & = & \lambda u_\mu u_\nu + \lambda \left( {\dot y}_\mu {\dot y}_\nu 
\pm u_\mu {\dot y}_\nu \pm u_\nu {\dot y}_\mu\right) + \alpha g_{\mu\nu}^\pm
~~~,
\label{pcGR3}
\end{eqnarray}
which can be rewritten such that its property as a, in general,
anisotropic fluid is obvious. For that, we refer to
\cite{Adv-HEP,PPNP}. 

When $T^\Lambda_{\pm~\mu\nu}$ is mapped to the real part, 
one obtains

\begin{eqnarray}
8\pi T_{\mu\nu}^\Lambda & = & \lambda u_\mu u_\nu + \lambda {\dot y}_\mu {\dot y}_\nu 
+ \alpha g_{\mu\nu}^\pm
~~~.
\label{pcGR3a}
\end{eqnarray}

As we will see further below, the $y_\mu$ can be written
as being proportional to the 4-velocity $u_\mu$. This allows us to write the (\ref{pcGR3a}) proportional to $\lambda$
as $\lambda \left( 1+ l^2A(r)\right) u_\mu u_\nu$ (here for
a central problem), where $l^2$ describes the correction 
due to the ${\dot y}_\mu {\dot y}_\nu$ 
term. The $A(r)$ can be very
large near the event-horizon, as will be seen later.

For the
$T_{00}$-component this implies pc-corrections
to the density of the dark energy proportional to $l^2$. This effect was already studied in
\cite{feoli}, where the effective potential deduced
there produces a repulsive potential near the event horizon.
 
Though, we will discuss
some solutions, at the end it will be more effective to
assume a {\it phenomenological approach}, with a specific
$r$-dependence of the dark energy density.

For the dark energy density we prefer to use a 
phenomenological
ansatz, namely

\beqa
\varrho_\Lambda & \sim & \frac{1}{r^{n+2}}
~~~.
\label{rholambda}
\eeqa
The value of $n$ has to be larger than 3, as discussed
earlier. The proportionality factor parametrizes the
coupling of the central mass to the dark energy.

After these general consideration, we resume the metric 
of a rotating star (Kerr metric), which was derived in 
\cite{hess2013}:

\beqa
g_{00} & = & -\frac{ r^2 - 2m_0 r  + a^2 \cos^2 \vartheta + \frac{B_n}{(n-1)(n-2)r^{n-2}} }{r^2 + a^2 \cos^2\vartheta}~~~, \nonumber \\
g_{11} & = & \frac{r^2 + a^2 \cos^2 \vartheta}{r^2 - 2m_0 r + a^2 +  \frac{B_n}{(n-1)(n-2)r^{n-1}}  }~~~, \nonumber \\
g_{22} & = &  r^2 + a^2 \cos^2 \vartheta~~~,  \nonumber \\
g_{33} & = & (r^2 +a^2 )\sin^2 \vartheta + \frac{a^2 \sin^4\vartheta \left(2m_0 r -  \frac{B_n}{(n-1)(n-2)r^{n-2}}  \right)}{r^2 + a^2 \cos^2 \vartheta}~~~,  \nonumber \\
g_{03} & = & \frac{-a \sin^2 \vartheta ~ 2m_0 r + a \frac{B_n}{(n-1)(n-2)r^{n-2}}   \sin^2 \vartheta }{r^2 + a^2 \cos^2\vartheta}
~~~.
\label{pcGR-10}
\eeqa
The $B_n$ and $a$ are the above mentioned coupling parameter
of the mass to the dark energy and the Kerr rotational
parameter, respectively. In what follows $n=4$ will be assumed and
(\ref{pcGR-10}) reduces to the standard Kerr solution of GR
when $B_n=0$.

For $n=4$ and $b_n=\frac{81}{8}$
there is still for $a=0$ an event horizon at
$\frac{3}{2}m_0$. Thus, if we want to eliminate completely
this horizon, the value of $b_n$ has to be at least
infinitesimal larger than $\frac{81}{8}$. For
practical reasons $b_n$ is assumed to acquire
the value $\frac{81}{8}$.
In that case, a mass function $m(r)$ can be defined,
namely

\beqa
m(r) & = & \left( 1- \frac{27}{32} \left(\frac{m_0}{r}
\right)^3 \right)
~~~.
\label{mass}
\eeqa
For the pc-Schwarzschild case the $g_{00}$ component
acquires the form $g_{00}$ =  
$\left( 1 - \frac{2m(r)}{r} \right)$, which clearly
reduces to the well known expression in GR,
when term proportional to $1/r^3$ in $m(r)$ is
eliminated.

\subsection{Group theoretical properties of the
pseudo-complex description}

A more complete explanation of the 
group theoretical structure of the pc-coordinates
and their importance in field theory can be found in 
\cite{book,schuller1,schuller2,schuller3,schuller4}.
 
The pc-Lorentz transformation is

\beqa
e^{-\omega^{\mu\nu}L_{\mu\nu}} & = &
e^{-\omega_+^{\mu\nu}L^+_{\mu\nu}}\sigma_+ + e^{-\omega^{\mu\nu}L^-_{\mu\nu}} \sigma_-
~~~,
\label{finite-pc-trafo}
\eeqa
where $L^\pm_{\mu\nu}$ is the generator of the Lorentz
group $SO_\pm (3,1)$, restricted to the $\pm$ zero-divisor
component. Its explicit form is given by ($\hbar = 1$)

\beqa
L_{\mu\nu} & = & X_\mu P_\nu - X_\nu P_\mu
\nonumber \\
& = & 
\left( X^+_\mu P^+_\nu - X^+_\nu P^+_\mu \right) \sigma_+
+ \left( X^-_\mu P^-_\nu - X^-_\nu P^-_\mu \right) \sigma_-
\nonumber \\
& = & L^+_{\mu\nu} \sigma_+ + L^-_{\mu\nu} \sigma_-
\nonumber \\
{\rm with} &&
\nonumber \\
P_\mu & = & \frac{1}{i}\frac{\partial}{\partial X^\mu} 
~~~.
\label{lorentzgen}
\eeqa

With this and $\left[ X^\mu , P_\nu \right]$
= $i\delta_{\mu\nu}$, the commutation relations
of the pc-Lorentz transformation generators
are \cite{book}

\beqa
\left[ L_{\mu\nu} , L_{\lambda\delta} \right]
& = & i\left( g_{\lambda\nu} L_{\delta\mu}
+g_{\delta\mu}L_{\lambda\nu} + g_{\lambda\mu}L_{\nu\delta}
+g_{\delta\nu}L_{\mu\lambda} \right)
~~~.
\label{Lo-1}
\eeqa 

Thus, the group structure is

\beqa
SO_{pc}(3,1) & = & SO_+(3,1) \otimes SO_-(3,1) ~\supset~ SO(3,1)
~~~,
\label{group-structure}
\eeqa
i.e., the direct product of two Lorentz groups. This direct
product reduces to the standard Lorentz group, when the real 
part of the pc-generators is projected. 

In conclusion, the mathematical structure is very similar to
the standard formulation, it only involves twice as much generators and parameters for the transformation,
a reflection of the 8-dimensional coordinate space.

\section{Observational predictions}
\label{obs}

In this section, some older, with additions, 
and some new predictions 
of the pc-GR are presented. 
First of all, the general properties of pc-GR 
related to the $y^\mu$-components are discussed. 
As mentions above, the $y_\mu$ variables
can be identified, for the case of a flat space, by $lu_\mu$,
where the $l$ is a length parameter introduced for 
dimensional reasons. An allowed ansatz for a general
case is

\beqa
y_\mu & = & A(x,y) lu_\mu
~~~,
\label{ansatzmu}
\eeqa
where $A(x,y)$ is a function in $x_\nu$ and $y_\nu$, yet
to be determined, where $y_\mu$ has 
the same tensorial properties
as $u_\mu$ and $A(x,y)$ is a scalar.
That (\ref{ansatzmu}) is correct will be seen
in two examples to be discussed in this paper,
namely for the pc-Schwarzschild solution and
for the pc-Robertson-Walker metric. 

\subsection{A general equation for $y_\mu$}

In what follows, the general structure is discussed and
an equation for the determination of (\ref{ansatzmu})
is presented. At the end, the $y_\mu$ is obtained within an
approximation for the Schwarzschild case, the first
example. 
 
A curve in the pseudo-complex manifold is described by
the pc-4-vector

\beqa
V^\mu ~=~ \frac{dX^\mu}{ds} & = & \frac{dX^\mu_+}{ds}\sigma_+ +  \frac{dX^\mu_-}{ds}\sigma_-
~~~,
\label{eq-1}
\eeqa
with "s" as the curve parameter, which is the same in both
zero-divisor components.

In addition to the pc-coordinates $X^\mu$ the
{\it local} coordinates (or co-moving frame) 
${\widetilde X}^\mu$ at a given point
{\it p} are defined in analogy as

\beqa
X_\pm^\mu & = & x^\mu \pm y^\mu ~,~
{\widetilde X}_\pm^\mu ~ = ~ 
{\widetilde x}^\mu \pm {\widetilde y}^\mu
~~~.
\label{eq-1a}
\eeqa

Transforming (\ref{eq-1}) into the co-moving frame in both components, we obtain

\beqa
\frac{dX^\mu}{ds} & = & e^\mu_{+~i}\frac{d{\tilde X}^i_+}{ds} \sigma_+ +
e^\mu_{-~i}\frac{d{\tilde X}^i_-}{ds} \sigma_-
~~~,
\label{eq-2}
\eeqa
where $e^i_{\pm~ \mu}$ ($i$=0,2,3,4) are the 
inverse
matrix elements of $e^\mu_{\pm~ i}$. In this contribution
only diagonal metrics are discussed, therefore
the $e^i_{\pm ~ \mu}$ are the square roots of the
diagonal metric elements.

In a co-moving frame the space is locally flat and, as
we saw earlier, the ${\tilde y}^\mu$ assume the relation
$l\frac{d{\tilde x}^\mu}{ds}$, i.e., it is proportional
to the 4-velocity. This results into a relation of the
zero divisor components of $X^\mu$ in the local frame
to $x^\mu$ and $y^\mu$:

\beqa
\frac{d{\tilde X}^i_+}{ds} & = & e^i_{+~\mu} \frac{dX^\mu_+}{ds}
~=~ \frac{d{\tilde x}^i}{ds} + \frac{d{\tilde y}^i}{ds} ~=~ 
\left( \frac{1}{l}\right){\tilde y}^i + \frac{d{\tilde y}^i}{ds}
\nonumber \\
\frac{d{\tilde X}^i_-}{ds} & = & e^i_{-~\mu} \frac{dX^\mu_-}{ds}
~=~ \frac{d{\tilde x}^i}{ds} - \frac{d{\tilde y}^i}{ds} ~=~ 
\left( \frac{1}{l}\right){\tilde y}^i - \frac{d{\tilde y}^i}{ds}
~~~.
\label{eq-3}
\eeqa

Next, we solve for $\widetilde{y}^\mu$ and its derivative
with respect to $s$, giving

\beqa
{\tilde y}^i & = & l
\left[
\frac{1}{2} e^i_{+~\mu} \frac{dX^\mu_+}{ds} + 
\frac{1}{2} e^i_{-~\mu} \frac{dX^\mu_-}{ds}
\right]
\nonumber \\
\frac{d{\tilde y}^i}{ds} & = & \frac{1}{2} 
e^i_{+~\mu} \frac{dX^\mu_+}{ds} -
\frac{1}{2} e^i_{-~\mu} \frac{dX^\mu_-}{ds}
~~~.
\label{eq-4}
\eeqa 
In a subsequent 
step, the first equation in (\ref{eq-4}) is derived further
with respect to the line element $s$ and the result is set equal to the second equation in (\ref{eq-4}).
Then, all expression of the $\sigma_+$ and 
$\sigma_-$ components are collected
on one side of the equation, resulting into

\beqa
&
l\frac{de^i_{- ~\mu}}{ds} 
\frac{d X^\mu_-}{ds} 
+ e^i_{-~\mu} \left[ l\frac{d^2 X^\mu_-}{ds^2} + \frac{dX^\mu_-}{ds} \right]
&
\nonumber \\
&
-e^i_{+~\mu}\left[ \frac{dX^\mu_+}{ds} - l\frac{d^2 X^\mu_+}{ds^2} \right]
+ l\frac{de^i_{+\mu}}{ds} \frac{dX^\mu_+}{ds} ~=~ 0
~~~.
&
\label{eq-6}
\eeqa

Let us consider, as a special case, a central and static 
problem: Set
$dX_\pm^0=0$ and $dX_\pm^\vartheta =0$,
$dX_\pm^\varphi =0$. As an approximation, 
all terms proportional to the minimal length  
are neglected compared to the others, because 
we will not work in the regime of maximal acceleration.
We also rescale, for this particular example, the mass parameter to $m_0=1$.

The 4-bein tensors $e^i_{\pm ~ \mu}$ are used, 
restricting to the pure $r$ part: 

\beqa
e_{\pm ~ \mu}^i & \rightarrow & e_{\pm ~ r}^r ~=~ 
1-\frac{2m_\pm (X_\pm^r )}{X_\pm^r}
~,~ m_\pm ~ = ~ 1- \frac{27}{32 X_\pm^r}
~~~.
\label{eq-7}
\eeqa
The differential equation
(\ref{eq-6}) then reduces to

\beqa
\frac{dy_r}{dr} & = &
\frac{
\left[
1+\frac{27}{16(r-y_r)^4} - \frac{2}{(r-y_r)}
\right]^\frac{1}{2}
-
\left[
1+\frac{27}{16(r+y_r)^4} - \frac{2}{(r+y_r)}
\right]^{\frac{1}{2}}
}
{
\left[
1+\frac{27}{16(r-y_r)^4} - \frac{2}{(r-y_r)}
\right]^\frac{1}{2}
+
\left[
1+\frac{27}{16(r+y_r)^4} - \frac{2}{(r+y_r)}
\right]^\frac{1}{2}
}
~~~.
\label{eq-8}
\eeqa

Furthermore, assuming a small $y_r$, 
compared to $r$, we arrive at

\beqa
\frac{dy_r}{dr} & \approx & \frac{(\frac{27}{8 r^5} - 
\frac{1}{r^2})}{1+\frac{27}{16 r^4} - \frac{2}{r}}
y_r
~~~.
\label{eq-9}
\eeqa
When the differentials $dy_r$ and $dr$ are moved to opposite 
sides of the equation and we divide it by $d\tau$, where $\tau$
is the time parameter, one notes that the time derivative
of $y_r$ is proportional to the corresponding component
of the four-velocity, as noted further above.

The MATHEMATICA code \cite{mat11} is used
to solve this equation, with the result

\beqa
y_r & = & A \frac{r^2}{(3-2 r)(3+4r+4r^2)^\frac
{1}{2}}
~~~.
\label{eq-10}
\eeqa
The integration constant $A$ is determined in the 
limit of $r \rightarrow \infty$. Requiring that
$y_r$ should approach $l{\dot r}_\infty$, leads to
$y_r  \rightarrow  -\frac{A}{4}$, or $A=-4l{\dot r}_\infty$.
In this manner, the
minimal length $l$ is reintroduced due to the 
assumed limiting behavior of $y_\mu$.

Thus, the final solution is

\beqa
y_r & = & \frac{4 l r^2}{(2r-3)(3+4r+4r^2)^\frac{1}{2}}
{\dot r}_\infty
~~~.
\label{eq-12}
\eeqa
This solution has a pole in $r=\frac{3}{2}$, where $y_r$
tends to $+\infty$ for $r \rightarrow (\frac{3}{2})\mid_+$. Because
we assumed $y_r$ as small compared to $r$, the solution is not
valid anymore  near $r=\frac{3}{2}$
and a solution for large $y_r$ has to be found.

\begin{figure}[t]
\begin{center}
\rotatebox{0}{\resizebox{180pt}{200pt}{\includegraphics[width=0.5\textwidth]{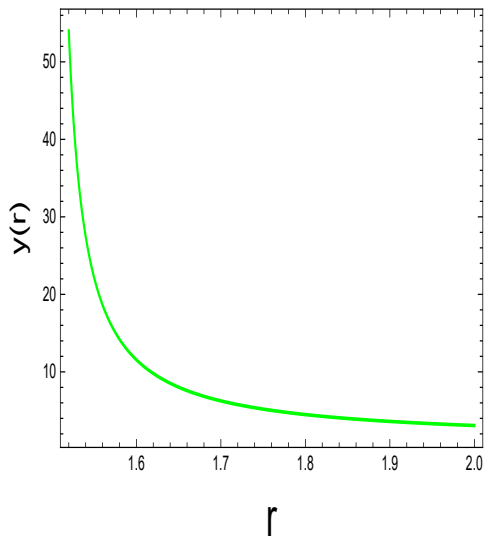}}}
\end{center}
\caption{
The radial pseudo-imaginary coordinate component
$y(r)=y_r/{\dot r}_\infty$ as a function in r. 
The vertical axis gives 
$y(r)$ in units of $l$.
\label{app-1}
}
\end{figure}

Nevertheless, the solution describes the evolution of $y_r$
from large $r$ towards smaller values, when $y_r$ is still small. The limit is when $(2r-3)$ in the denominator
is of the order of one, or $r  \approx  
\frac{3}{2} + \frac{l}{2}$, i.e., {\it very close} to $r=\frac{3}{2}$. A general 
picture emerges, where $y_r$ is slowly increasing for 
descending  $r$-values and changes rapidly toward 
$+\infty$ near $r=\frac{3}{2}$. 
In Fig. \ref{app-1} the function 
$y_r/{\dot r}_\infty$, in units of 
the minimal length $l$,
is plotted in the range 1.52 to 2 and, as can be seen, 
only near 1.5 the curve starts to rise.

\begin{figure}[ht]
\begin{center}
\rotatebox{270}{\resizebox{180pt}{220pt}{\includegraphics[width=0.23\textwidth]{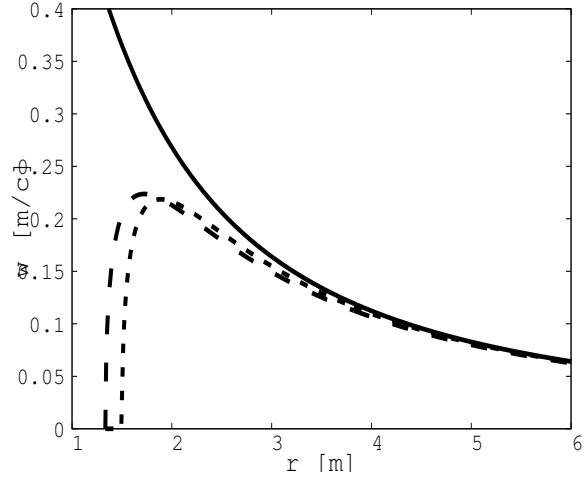}}}
\end{center}
\caption{ 
The angular frequency of a particle in a circular orbit,
as a function in the radial distance and for $a=0.95m_0$.
The upper curve is the result as obtained within GR and the
two lower curves within pc-GR. The curve with the maximum
to the left is for $n=3$ and the other one for $n=4$.
}
 \label{fig1}
\end{figure}

\begin{center}
\begin{figure}[ht]
\begin{center}
\rotatebox{0}{\resizebox{240pt}{160pt}{\includegraphics[width=0.23\textwidth]{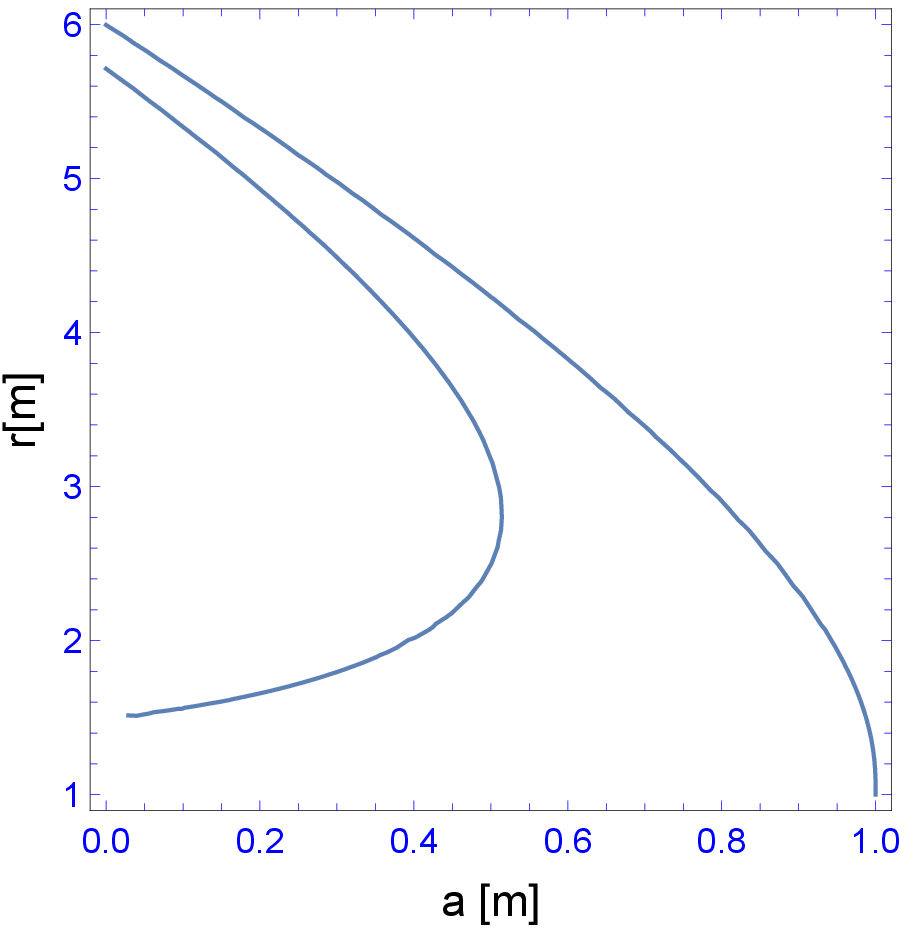}}}
\end{center}
\caption{ 
Limits of stable orbits within GR and pc-GR ($n=4$). 
The upper curve is for GR, and the inner curve for pc-GR.
Below the upper curve, no stable orbits exist, while
in pc-GR no stable orbits exist to the left of the inner curve.
Thus, in pc-GR, for $a$-values to the right of the inner
curve all orbits are stable.
}
 \label{fig2}
\end{figure}
\end{center}

\subsection{Perihelion shift of Mercury}
\label{mercury}

The first confirmation of General Relativity came from the
calculation of the perihelion shift of Mercury. 
As an illustrative example, in this 
sub-section, the perihelion shift is investigated, 
including the pc-corrections, with the modified mass-function
$m(r)$ = $m_0\left(1-\frac{27}{32}\frac{m_0^3}{r^3}\right)$
(now, we return to $m_0 \neq 1$),
where the paths as described in \cite{adler}
is neatly followed.
We will see, that GR gives the dominant contribution,
while the additional ones from pc-GR will be too low to be
measured.
That this correction cannot be measured
is not surprising, because the 
gravitational
field in the solar system is simply too weak. The
consequences for all other solar system observations
are equally small. 

The real length element in the pc-Schwarzschild 
solution is

\beqa
ds^2 & = & -\left( 1 - \frac{2m(r)}{r} \right) dt^2
+ \left( 1 - \frac{2m(r)}{r} \right)^{-1}dr^2
\nonumber \\
&&
+r^2 \left(d\theta^2 + {\rm sin}^2\theta d\varphi^2\right)
~~~,
\label{merc1}
\eeqa
divide it by $ds^2$ and use for the Lagrangian the
definition $L=-1$ \cite{hess2013}, which leads to

\beqa
1 & = & \left( 1 - \frac{2m(r)}{r} \right) {\dot t}^2
- \left( 1 - \frac{2m(r)}{r} \right)^{-1}{\dot r}^2
\nonumber \\
&&
-r^2 \left({\dot \theta}^2 + {\rm sin}^2\theta 
{\dot \varphi}^2\right)
~~~,
\label{merc2}
\eeqa
where the dot refers to the derivative in $s$.
Using the angular momentum conservation $l=r^2 \dot{\varphi}$
= ${\rm const}$, that the motion is in the plane of
$\theta = \frac{\pi}{2}$ and that
$\left( 1-\frac{2m(r)}{r}\right){\dot t}$ = $h$ = 
${\rm const}$ 
(which is the result of of the variation with respect 
to the time), we arrive at 

\beqa
1 & = & \left( 1 - \frac{2m(r)}{r} \right)^{-1} h^2
- \left( 1 - \frac{2m(r)}{r} \right)^{-1}{\dot r}^2
- \frac{l^2}{r^2}
~~~.
\label{merc3}
\eeqa

Changing the derivative with respect to $s$ to the one 
with respect to $\varphi$, leads to

\beqa
r^\prime & = & \frac{dr}{d\varphi} ~=~
\frac{\dot r}{\dot \varphi}
~ \rightarrow ~ {\dot r} ~=~ {\dot \varphi}r^\prime ~=~
\frac{l}{r^2} r^\prime
~~~,
\label{merc4}
\eeqa
where the prime now refers to the derivative with respect
to $\varphi$.

Multiplying (\ref{merc3}) by $\left( 1 - \frac{2m(r)}{r} \right)$ and using (\ref{merc4}), we arrive at the
equation

\beqa
\left( 1 - \frac{2m(r)}{r} \right) & = &
h^2-\frac{l^2}{r^4}(r^\prime )^2 
- \frac{l^2}{r^2}\left( 1 - \frac{2m(r)}{r} \right)
~~~.
\label{merc5}
\eeqa

Next, the variable $r$ is changed to $u$, namely

\beqa
r & = & \frac{1}{u} ~\rightarrow~ r^\prime ~=~
-\frac{u^\prime}{u^2}
~~~.
\label{merc6}
\eeqa

Substituting this into (\ref{merc5}) and resolving for
$(u^{\prime})^2$, one arrives at

\beqa
(u^{\prime})^2 & = & 
\frac{(h^2-1)}{l^2} +\frac{2m(u)}{l^2}u
- u^2 + 2m(u)u^3
~~~,
\label{merc7}
\eeqa
where $m(u)$ is given by

\beqa
m(u) & = & m_0\left( 1-\frac{27}{32}m_0^3 u^3\right)
~~~.
\label{merc8}
\eeqa

In the next steps, (\ref{merc7}) is again derived
with respect to $\varphi$ and the resulting 
equation is divided 
by $2u^\prime$, arriving finally at the equation

\beqa
u^{\prime\prime} + u & = & \frac{m_0}{l^2} + 3m_0 u^2
-\frac{27}{8}\frac{m_0^4}{l^2} u^3
- \frac{81}{16} m_0^4 u^5
~~~.
\label{merc9}
\eeqa
Define the small number 
$\varepsilon = 3m_0 A$, which is for the solar system
of the order of $10^{-7}$ \cite{adler}, where 
$A=\frac{m_0}{l^2}$ is the
areal velocity on the planetary plane.
With this, the $m_0$
can be set equal to $\frac{\varepsilon}{3A}$ and
(\ref{merc9}) can be rewritten as

\beqa
u^{\prime\prime} + u & = & A + \frac{\varepsilon}{A} u^2
\nonumber \\
&&
-\frac{A}{8}\left(\frac{\varepsilon}{A}\right)^3u^3
- \frac{1}{16} \left(\frac{\varepsilon}{A}\right)^4 u^5 
~~~.
\label{merc9a}
\eeqa

In the first line, the first term on the right hand side is the classical, Newtonian
contribution, while the second term is the standard
relativistic correction which leads to the 
known perihelion shift of Mercury. 
The terms in the second row are explicit 
new contributions from pc-GR, however of order
$\varepsilon^3$. Thus, new contribution of pc-GR
are at least two order of magnitude less, the
one of order $\varepsilon^2$ coming only from GR
(implicitly contained in the term 
$\frac{\varepsilon}{A}u^2$ when $u$ is also expanded 
in $\varepsilon$). Thus,
there is no hope to detect these deviations.
This result also shows that GR and pc-GR are indistinguishable
from each other in solar system observations. Still, this
calculation is an interesting exercise, which might be
of relevance when the perihelion shift of objects very
near to a super-massive black hole is observed.

\subsection{Quasi Periodic Objects}
\label{QPO}

We mention shortly a former discussion on
so-called {\it Quasi-Periodic objects}. There are several kinds, but here we refer to bright light emissions in accretions discs with a near periodic time dependence.
At first, this phenomenon was associated to a bright
spot which is co-moving with the accretion disc, thus, 
appearing and fading away from the observer as it turns
around the black hole. Such events were observed
\cite{book,ws-book,QPO1,QPO2,QPO3,QPO4} 
in black hole binaries, but also exist in central black holes of galaxies. The GR can provide an estimation
their distance to the center of the black hole, 
using
the dependence in $r$ of the orbital period 
of a circular particle. 
In order to confirm this 
distance, the redshift of the emission line has 
to be measured, too, which depends on $r$. 
The radial distance deduced from the orbital frequency
and the redshift have to coincide for consistency.
For central
black holes the redshift has not been observed yet, but
it is for stellar black holes.

The assumption is that the QPOs are the result of a bright 
spot moving around the black hole, which is obvious for
black holes in the center of galaxies. However, for
stellar black holes there may be a distinct mechanisms, 
like
oscillations provoked by the companion star, as suggested
in \cite{QPO-GR}. Thus, GR can still be saved. The
question remains: Why the interpretation of OPOs should be 
different from the ones in the accretion disc of a  
black holes in the center of a galaxy? We argue that
the same
mechanism should also hold for stellar black holes,
at least dominantly.

In order to understand the motion of a point-particle
around a black hole, we resume
some fundamental properties of this particle in a circular
orbit: In Fig. \ref{fig1} the dependence of the orbital
frequency as a function in $r$ is depicted for the
rotational Kerr-parameter $a=0.95m_0$. The upper curve is
the result for GR, while in the lower curve the one for
pc-GR is plotted. In GR the curve shows a 
steady increase of the orbital frequency, contrary to
pc-GR which predicts a maximum, after which if falls off again. 

\begin{figure}[ht]
%\begin{center}
\rotatebox{0}{\resizebox{200pt}{200pt}{\includegraphics[width=0.5\textwidth]{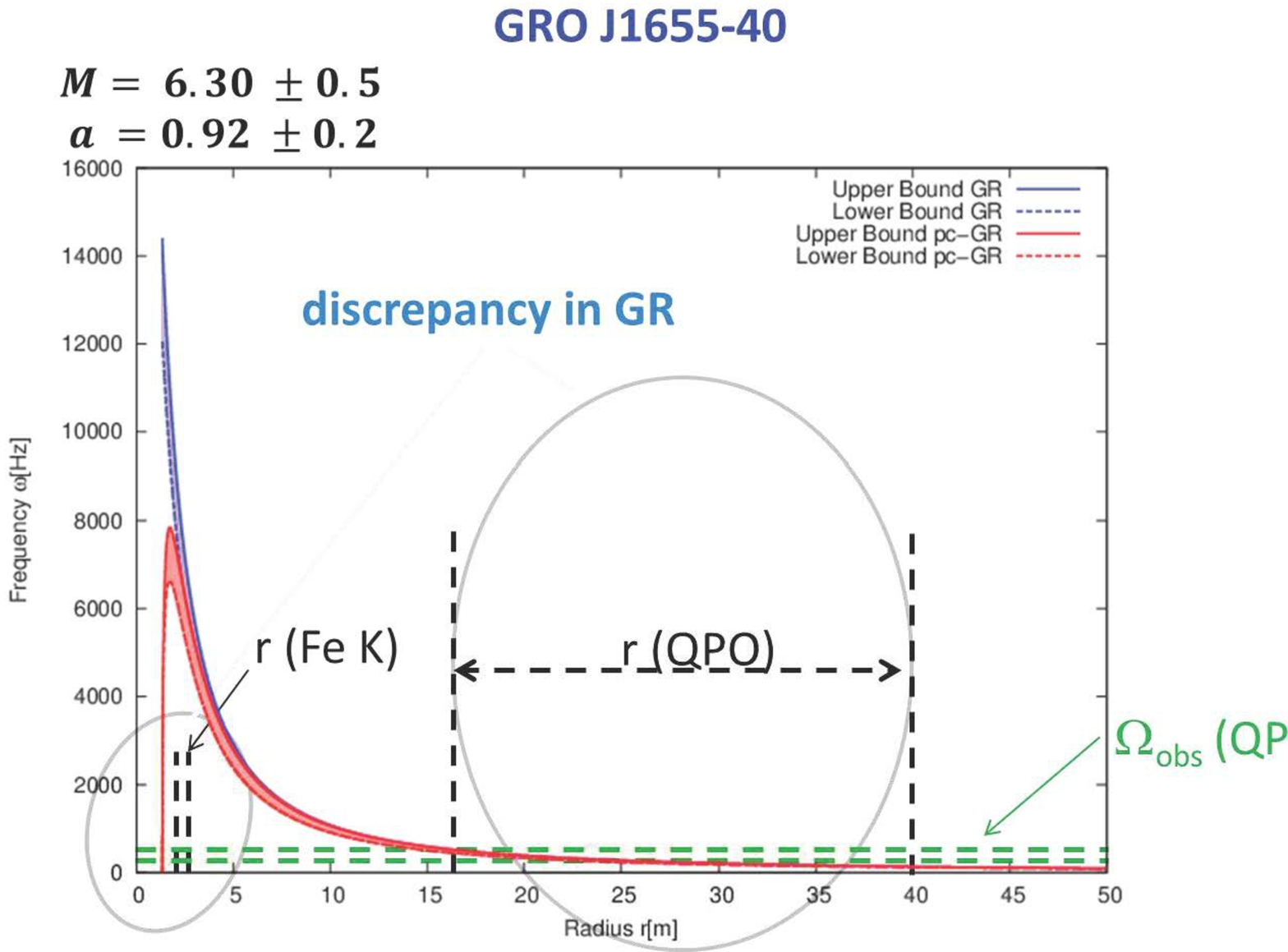}}}
\rotatebox{0}{\resizebox{200pt}{200pt}{\includegraphics[width=0.5\textwidth]{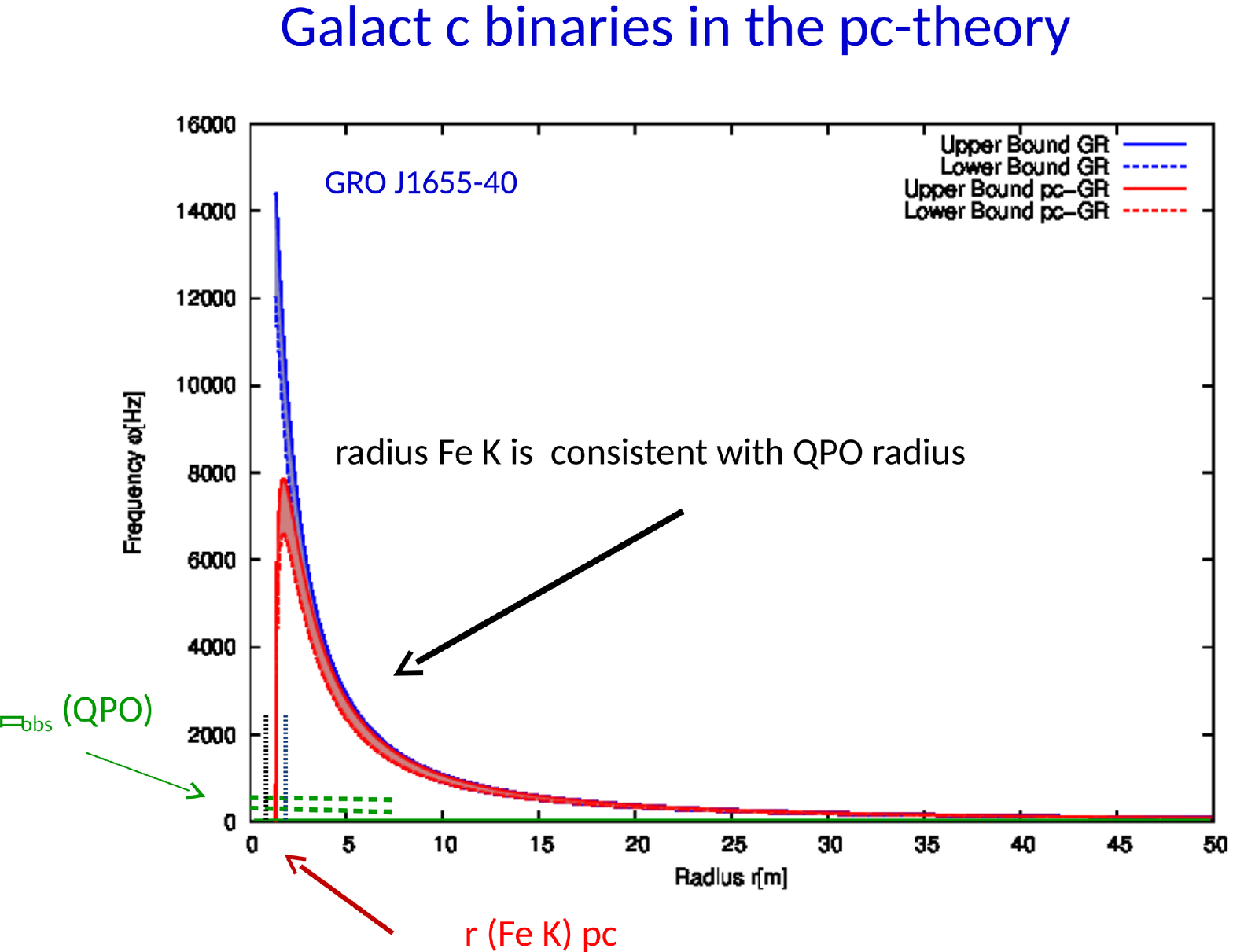}}}
%\end{center}
\caption{
{\it Quasi-Periodic Objects} (QPO) compared
to GR and pc-GR. The upper curve corresponds to the orbital frequency deduced within GR, where the thickness of the line
indicates the resolution of the observation. 
The lower curve corresponds to pc-GR for $n=3$. For $n=4$ 
the differences are minor, which is the reason
no new calculations were performed. 
The horizontal dashed lines show the
observed orbital frequency with the experimental error,
while the vertical lines show the deduced radial distance 
from the redshift of the $K\alpha$-line. 
\label{pcGR-fig5}}
\end{figure}

Deducing a distance from the orbital period and from the
redshift, leads for $n=3$ to Fig. 
\ref{pcGR-fig5}, where the observation is compared to the theory, using the $a$-parameter
deduced. 
The zero is for $n=3$ at $r=\frac{4}{3}m_0$
and for $n=4$ to $r=\frac{3}{2}$, which are 
practically indistinguishable within the plot, i.e.,
the consequences are the same.

There is a great mismatch between the 
distance deduced from the orbital frequency and the redshift,
using GR. In pc-GR, however, the results agree. A simple
explanations leads to an agreement to observation!
Nevertheless, because GR can be still reconciled, though, 
via a more complex explanation, nothing can be decided
up to now.

\subsection{Accretion disc structure}
\label{disc}

In order to understand the predicted light emission
from an accretion disc in pc-GR, 
stable circular orbits have to be discussed: 
In Fig. \ref{fig2} the last stable orbit
(or ISCO for {\it Innermost Stable Circular Orbit}) 
as a function in $a$ is shown. In GR the ISCO
starts at $6m_0$ and is lowered 
to $r=m_0$ for $a=1$.
Below the upper line in Fig. \ref{fig2} no stable
orbit exists. This is different in pc-GR: The
lower curve shows the limits of stable orbits, where to
its left no stable orbit exists but to its right it does.
The curve turns at about $a=0.5m_0$, thus above
$a=0.5m_0$ all orbits are stable. It is this region
where they can reach the maximum of the orbital frequency,
below  this value of $a$ they don't. 

For small $a$, the curve for pc-GR 
follows neatly the one of GR.
Thus, we do expect a similar emission profile structure, 
except for the intensity of the 
light emitted, because the curve is further
into the gravitational well and more energy is released.
When $a>0.5m_0$ the stable orbits reach the maximum of the
orbital frequency. At this maximum, neighboring orbitals have
a similar frequency and the excitation of the disc is
minimal, producing a dark ring. To resume, pc-GR provides
a robust prediction of the
emission profile with a dark ring follows further in
by a small bright ring. 

\begin{figure}[t]
%\centerline{
\begin{center}
\rotatebox{0}{\resizebox{130pt}{130pt}{\includegraphics[width=0.5\textwidth]{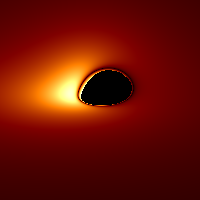}}}
\rotatebox{0}{\resizebox{130pt}{130pt}{\includegraphics[width=0.5\textwidth]{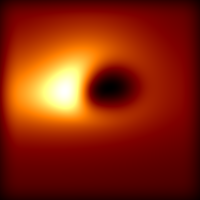}}}
\end{center}
%\begin{center}
%\rotatebox{270}{\resizebox{150pt}{150pt}{\includegraphics[width=0.5\textwidth]{M87-EHT.jpg}}}
%\end{center}
%}
\caption{
The simulations on the left and right panel are within
pc-GR. The resolution on the left is for $5\mu$as while
on the right it is for $20\mu$as. The inclination angle
with respect to the observer is $70^o$. Clearly seen is
that the ring structure, as predicted by pc-GR, is
washed out, showing that the EHT observation cannot
discriminate between pc-GR and GR.
\label{pcGR-fig3}}
\end{figure}

The structure of the emission on accretion discs is
discussed in \cite{boller1,boller2}, comparing GR with
pc-GR, besides what happens when a cloud is approaching
the black hole.

In Fig. \ref{pcGR-fig3} a simulation is shown, using the model
published in \cite{page}, which assumes a thin,
optical thick accretion
disc. This is not a very realistic 
model, expecting rather a thick disc,
maybe with the form of a torus. For such
discs, other models are better suited, from which we just
mention one, namely \cite{kluzniak}. Nevertheless,
the emission structure is robust
and should be similar in all disc models, changing only the
absolute value of the intensity emitted.

On the left hand side of Fig. \ref{pcGR-fig3} the result for a  high solution is depicted, while the right hand side
shows the result for a low resolution of $20\mu$as.
For the low resolution case, the ring structure is 
unfortunately lost. The {\it Event Horizon Telescope}
(EHT) \cite{EHT1,EHT2,EHT3,EHT4,EHT5,EHT6}, which
observed the shadow of the central black hole of M87, has a resolution of
only $24\mu$as, which is too low to dissolve the ring 
structure. Therefore, one has to wait for future
observations with an improved resolution.
Note that the EHT is not able to distinguish between GR
and different kinds of extensions

The disc model used by EHT consists of an optical thin plasma
surrounding the black hole (see also \cite{johnson2020}. 
This is only one of many
possible disc models, most of which are optical thick.
The shadow is simulated tracing the most inner photon 
(or Einstein) rings. While in GR for $a=1m_0$
all photon rings are blocked by a optical
thick accretion disc, because the ISCO reaches $r=m_0$,
in pc-GR all photon rings are blocked, because for
$a>0.5m_0$ all orbits are stable until to the surface
of the star. A definite answer can only be given, when the
disc is proven to be optical thin and/or pc-GR is confirmed 
or not.

\subsection{Cosmology: The Robertson-Walker Model}
\label{RWM}

The pc-Robertson-Walker metric was introduced in
\cite{mag-2010} and can also be retrieved form
\cite{book}. The length element is given by

\beqa
d\omega^2 & = & (dX^0)^2 - \frac{a(t)^2}
{\left( 1+\frac{ka(t)^2}{4a(0)^2} \right)^2} d\Sigma^2
~~~,
\label{rel-2}
\eeqa
where $d\Sigma^2$ is the angular volume element, $a(t)$
is the radius of the universe, $a(0)$ its value at
present time (set to 1) and $k$ is a parameter, which only
for $k=0$ corresponds to a flat universe \cite{adler,misner}.
This value will be used from here on, because all observational
date suggest a flat universe.

For the energy-momentum tensor an isotropic fluid is
assumed, i.e.,

\beqa
\left( T^\mu_\nu \right) & = &
\left(
\begin{array}{cccc}
\rho &&& \\
& -\frac{p}{c^2} && \\
&& -\frac{p}{c^2} & \\
&&& -\frac{p}{c^2}  \\
\end{array}
\right) ~~~,
\label{energy-momentum}
\eeqa

Using the metric (\ref{rel-2}), the
$y_\mu$ can be determined, 
using the differential equation
(\ref{eq-6}). Because an isotropic universe is assumed,
the development of the radius $a(t)$ is the same
in any direction. Therefore, it suffices to define
$y_\mu$ as $a_I(t)$.

In analogy to (\ref{eq-8}) and using
$a(t)=a_R(t)+Ia_I(t)$, we arrive at

\beqa
\frac{da_I}{da_R} & = & \frac{
\left[ \frac{(a_R(t)-a_I(t))^2}{a(0)^2} 
- \frac{(a_R(t)+a_I(t))^2}{a(0)^2}
\right]
}{
\left[ (\frac{a_R(t)-a_I(t))^2}{a(0)^2} 
+ \frac{(a_R(t)+a_I(t))^2}{a(0)^2}
\right]
}
\nonumber \\
& = & -4\frac{a_R a_I}{a_R^2 + a_I^2}
~~~.
\label{rel-3}
\eeqa 

Because the $a_I$ has to be very small compared
to $a_R$, we obtain approximately, setting $a(0)=1$,

\beqa
\frac{d a_I}{d a_R} & = & -4 \frac{a_I}{a_R}
~~~.
\label{rel-4}
\eeqa
Separating variables and integrating, we finally obtain

\beqa
a_I & = & A\frac{1}{a_R^4}
~~~.
\label{rel-5}
\eeqa
Demanding an $a_I$ proportional to $l{\dot a}_R(0)$ 
at $t=0$, 
as similarly done in the Schwarzschild case, we can 
set $A=l{\dot a}_R(0)$. Thus, today $a_I(0)=l{\dot a}_R(0)$
(compared to $a_R(0)=1$), which is extremely small,
and it continues to be even smaller for larger times.
However, for $t \rightarrow 0$ (time of the Big Bang),
the $a_I$ explodes and becomes infinite
very near to the big bang, 
implying a dominance
of the dark energy. Of course, in this case, the 
approximation of a small $a_I$, compared to $a_R$ is not 
valid anymore, but the results enlightens the 
{\it tendency} of a strong increase of $a_I(t)$. 
Apart from that, very near to the big bang the
mass/energy density is extremely high and 
according to our assumption additional 
vacuum fluctuations building up (see section \ref{neutron}).

In \cite{book,mag-2010} the possible fates of the universe
were determined. The equations of motion were set up and
the differential equation for the second derivative of
$a_R$ became, assuming that the matter distribution in the
universe behaves as dust,

\beqa
\frac{a_R^{\prime\prime}}{\frac{4\pi G}{3}} 
& = & (3 \beta -1) \Lambda 
a_R^{3(\beta - 1)+1} - \varrho_0
a_R^{-2}
~~~,
\label{rel-6}
\eeqa
where $\varrho_0$ is the matter density at the present date.
The $\beta$ is a parameter of the dark energy density
$\varrho_\Lambda = \Lambda a_R^{3(\beta -1)}$. For $\beta=1$
this energy density is just $\Lambda$, as it seems to be 
satisfied today. $G$ is the 
gravitational constant.

With this, the differential equation of the radius of
the universe is

\beqa
\frac{a_R^{\prime\prime}}{\frac{4\pi G}{3}\varrho_0}
& = & 2\Lambda a_R - a_R^{-2}
~~~.
\label{rel-7}
\eeqa
The acceleration starts with a negative sign in the early 
epoch of the universe until $a_R^{\prime\prime}=0$,
which is reached when $a_R= 1/(2\Lambda)^\frac{1}{2}$,
after which the acceleration becomes positive, as observed 
today.

In \cite{book,mag-2010} also other scenarios were discussed,
taking different values of $\beta$. As a result, besides
a rip-off for values $\beta > 1$, also solutions were found,
which for $t \rightarrow \infty$ approaches a
constant value or even a zero positive acceleration ($0_+$).

\subsection{Neutron stars}
\label{neutron}

\begin{figure}[ht]
\begin{center}
\rotatebox{0}{\resizebox{300pt}{250pt}{\includegraphics[width=0.23\textwidth]{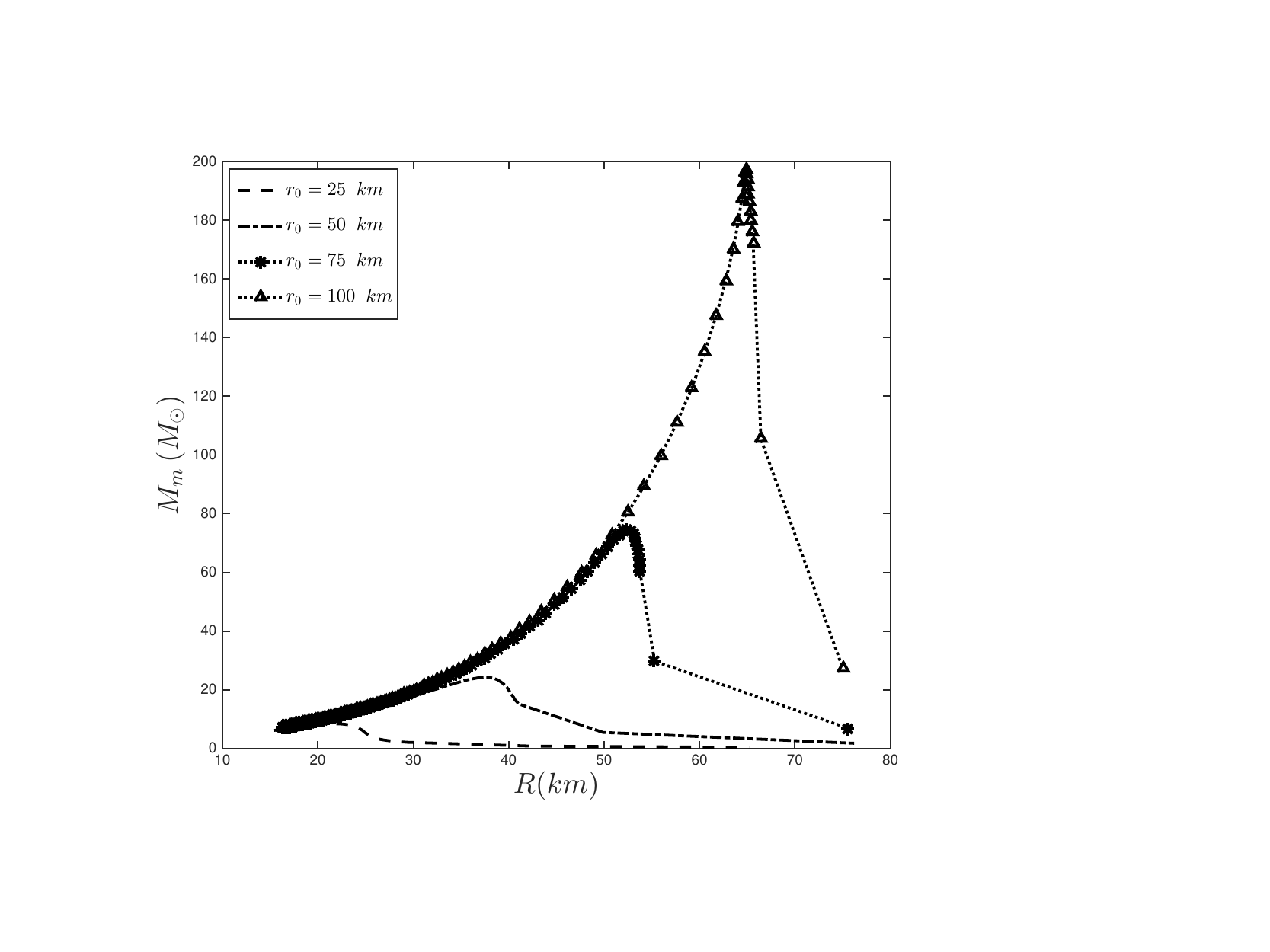}}}
\end{center}
\caption{ Within pc-GR, stable
stars up to 200 solar masses are obtained for a
coupling of the dark energy density to the mass density,
which diminishes approaching the surface.
}
 \label{200}
\end{figure}

\begin{figure}[ht]
\begin{center}
\rotatebox{0}{\resizebox{250pt}{160pt}{\includegraphics[width=0.23\textwidth]{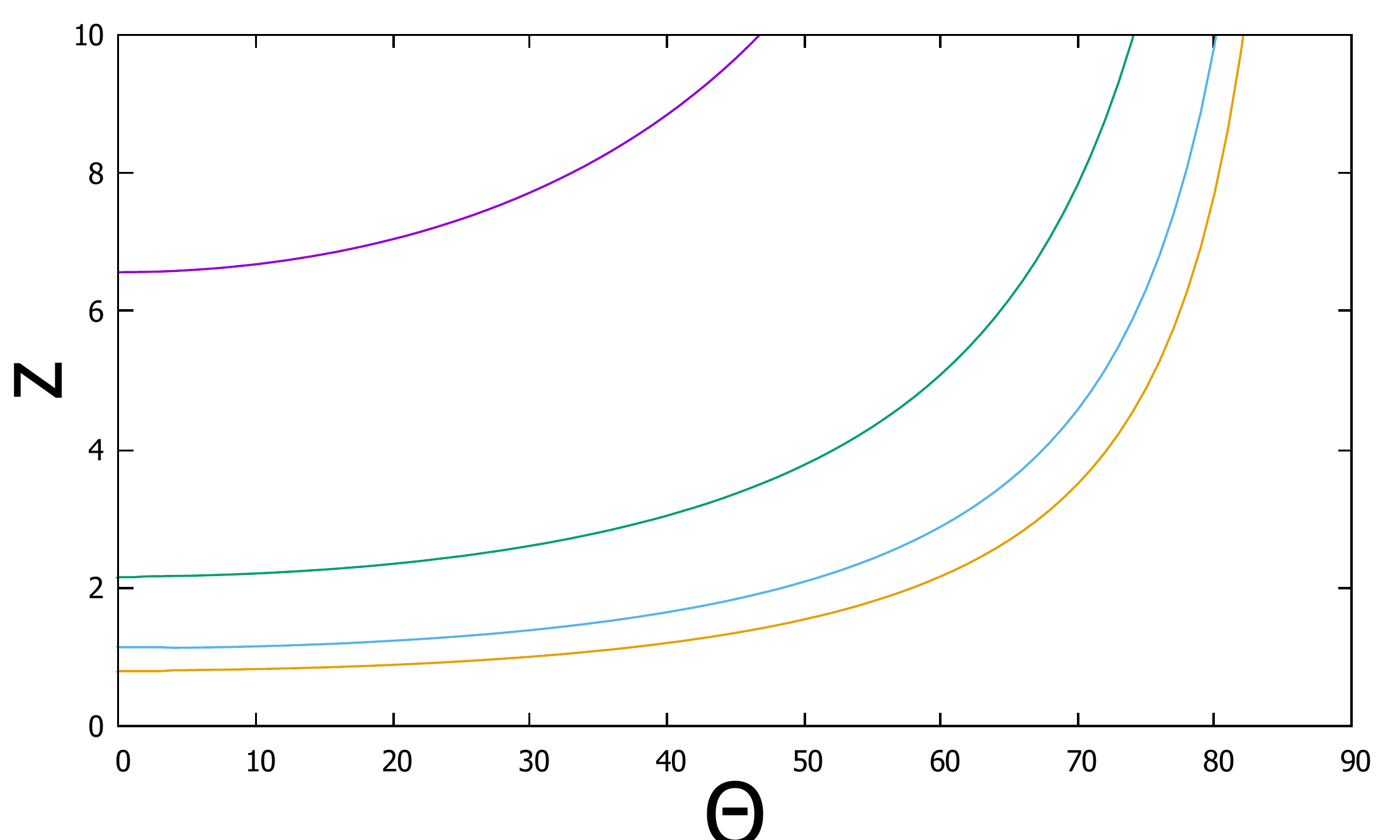}}}
\end{center}
\caption{The redshift at the surface $r=\frac{3}{2}m_0$
as a function of the azimuthal angle $\theta$, for
different $a$-values. From top to bottom, the curves
correspond to 0.2$m_0$, 0.5$m_0$, 0.8$m_0$ 
to 1.0$m_0$.
}
 \label{fig5}
\end{figure}

The pc-GR conjectures that there is no event horizon, 
implying that all so-called black holes are rather
gray stars, even when they have a mass billion times 
more than the sun. The question is, if one can
describe the internal structure of such stars and explain
their stability. One problem is that there is no such 
theory, because no one can deduce the equation of state
under such extreme mass densities. However, one can 
try to extrapolate from low masses to larger ones, using
recently developed theories for neutron stars.
Such a theory is published in \cite{schramm}, which
is a mean-field approach involving different hadron states
including strange hadron fields.

In \cite{rod-g41-2014,rod-tesis} this mean-field model
was coupled to dark energy within the star. As a 
first approximation, a linear coupling between 
the dark energy density to the mass density was assumed:

\beqa
\varrho_\Lambda & = & \alpha \varrho_m
~~~,
\label{neu1}
\eeqa
with the result of
up to 6 solar masses 
already found to be stable.
This is interesting in view of recent attempts to
explain higher neutron masses \cite{highmneut} due
to the possible detection of a neutron star with 2.6 solar
masses in a gravitational wave event \cite{neut-highmass}.
{\it This is a hint that even larger neutron
star masses can exist}, though, it should be expected
that their internal structure has to change significantly.

No higher masses could be obtained, because the linear
coupling resulted in a repulsion effect shedding off 
mass from the star for increasing mass. This was the
motivation to reanalyze vacuum fluctuations within the star,
using semi-classical Quantum Mechanics, as explained in
\cite{birrell}. In \cite{caspar-2016} the monopole
approximation \cite{monopole} was used and
the coupling between the dark energy and mass density
deduced. As a result a decrease of the coupling near the 
surface was obtained and stable solutions of 
up to 200 solar masses were obtained, see Fig, \ref{200}. 
This presented at
the same time a limit due to the
end of validity of the
mean-field theory \cite{schramm}. For higher masses other,
not yet developed theories have to be searched for.

How one could detect such large neutron stars (if one
can still denote them as neutron stars is a big question,
too)? One possibility is to look for emissions from
infalling matter and/or beams emitted as in standard neutron
stars. This is not as easy, because near the position of the
so-called event horizon the redshift is 
extremely large,
at least in the orbital plane. This is illustrated in
Fig, \ref{fig5}, where the redshift at $r=\frac{3}{2}m_0$
(which is the position of the event horizon for $a=0$, with the mass-function used)
is plotted versus the azimuthal angle $\theta$, for a set
of Kerr-rotational parameters $a$. As can be 
seen, near the poles
the redshift is, for large values of
$a$, of the order of 1. Thus, if $a$ 
is near to 1, highly redshifted  light emission may be seen near
the poles, when matter falls in there. When, however,
the so-called black hole has an accretion disc, there 
has to be a jet
emerging near the poles, over-shining this effect.
Hopefully, one can observe it in SgrA* which is believed
to be devoid of an accretion disc. 

It remains to be mentioned that in \cite{volkmer1,volkmer2}
also neutron stars, as compact objects, were investigated
within pc-GR from the viewpoint of an effective field
theory. In \cite{volkmer3,volkmer4} the effects of dark matter
within pc-GR were also investigated. All these contributions
reflect a range of observable consequences and
applications of pc-GR for
neutron stars.

\subsection{Gravitational waves}
\label{gw}

In 2016 the first observed gravitational wave event was 
reported in \cite{abbot1,abbot2}. Assuming that GR is the 
theory to describe the merger, a two-point approximation
seems to be justified, mainly because the ISCO is still 
far away from contact. The steps for obtaining the
{\it chirping mass} ${\cal M}_c$, using the two-point
approximation, are explained in
\cite{maggiore}. In \cite{hess-2016} the same approximation
was used, with the caveat that the two black holes can
approach each other until their event-horizon touch each 
other, which renders the two-point approximation 
senseless. Nevertheless, this approach serves to extract
trends on how the deduced chirping mass changes.
The modified equation for the chirping mass is

\beqa
{\cal M}_c ~=~
{\widetilde {\cal M}}_c F_{\omega}({\bar r})
& = &
\frac{c^3}{G}\left[ \frac{5}{96\pi^\frac{8}{3}} \frac{df_{{\rm gw}}}{dt} 
f_{{\rm gw}}^{-\frac{11}{3}}
\right]^{\frac{3}{5}}
~~~,
\label{frequ-gw}
\eeqa
where on the right hand side values of
the observed frequency and its change in time appears. The
left hand side contains the 
modified chirping mass
$\widetilde{{\cal M}}_c$ and the apparent chirping
mass ${\cal M}_c$, the number deduced in 
\cite{abbot1,abbot2}. Both are related by a factor
$F(R)$, where $R$ is the relative distance of the two
black holes.  In the calculation it is assumed that 
the mass of both is approximately equal as it was 
in the observed case. $n=3$ was used, for which 
the function 
is given by $F_{\omega}(R)$ = 
$\left(1 - \left(\frac{2R_S}{3R}\right)^2\right)$,
where $R_S=2m_0$ is the Schwarzschild radius of
one of the black holes. For $n=4$ the new expression is
$F_{\omega}(R)$ = 
$\left( 1-\left(
\frac{3R_S}{4R}\right)^3 
\right)$. 
In any case, the $F(R)$ tends to a small value until the 
touching configuration (twice the Schwarzschild radius
for equal mass companions), implying that the real
chirping mass is larger then the deduced one using
GR- Here it is important to note that the approach to deduce 
the mass in \cite{abbot1,abbot2} is {\it theory dependent},
as it is also in pc-GR. In pc-GR a larger mass should result
and, thus, also the luminosity distance has to be larger,
such that the same is observed on Earth.
How large, depends very much on the model used to describe
the inspiral phase. A simple two-point approximation does not
work and one needs more sophisticated methods, as
for example
a numerical relativistic hydrodynamical approach
\cite{rezzolla-book}.

In contrast,  
it is much easier to describe the ring-down modes of 
a black hole, i.e., after the two black holes have merged.
The ring-down modes are obtained investigating the
stability of the final black hole under metric perturbations.
For the Schwarzschild case,
this part is well described in the book by S. Chandrasekhar
\cite{chandra}, which can be 
directly extended to the pc-GR metric.
There are two types of modes, the negative parity
solutions, also called Regge-Wheeler modes 
\cite{regge-wheeler}, and the positive parity modes,
also called Zerilli modes \cite{zerilli}.
In \cite{hess-lopez-2019} the axial 
modes where calculated
for $n=3$ and in \cite{hess-lopez-2020} also the axial
modes were calculated, now for $n=4$.
The time dependence of the ring-down modes is
$e^{-i\omega t}$ = $e^{-i\omega_R t} e^{\omega_I t}$,
with $\omega = \omega_R + i \omega_I$, separated in
its real and imaginary part. For damped modes
the $-\omega_I$ has to be positive. 

\begin{figure}[ht]
\rotatebox{0}{\resizebox{120pt}{260pt}{\includegraphics[width=0.23\textwidth]{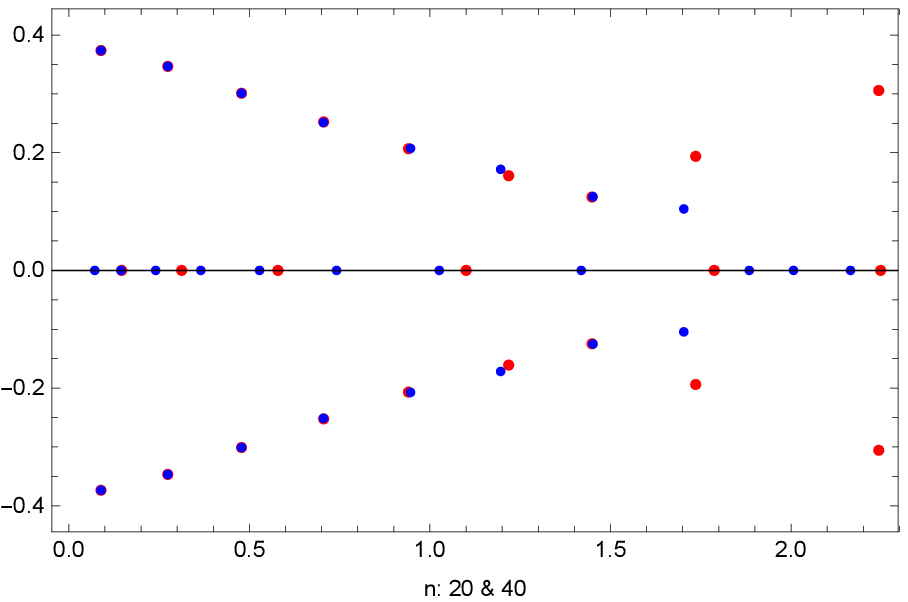}}}
\rotatebox{0}{\resizebox{120pt}{260pt}{\includegraphics[width=0.23\textwidth]{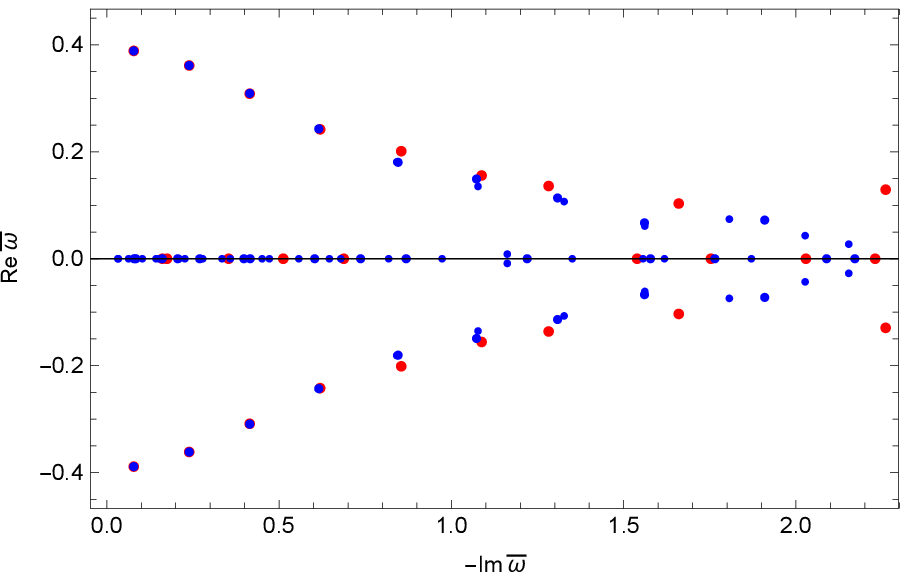}}}
\caption{
Axial gravitational modes within GR (left panel)
and pc-GR (within pc-GR). The iteration number
is 20 (red dots) and 40 (blue dots), using the
{\it Asymptotic Iteration Method} \cite{cho2012}.
}
 \label{axial}
\end{figure}

The Regge-Wheeler equation is solved as explained in 
\cite{hess-lopez-2019,hess-lopez-2020},
using an iterative technique called the
{\it Asymptotic Iteration Method} (AIM) 
\cite{cho2012}, and in 
Fig. \ref{axial} the result for small $-\tilde{\omega}_I$
is depicted (we defined $\tilde{\omega}=m_0\omega$). 
Note that there are no unstable modes with negative
$-\tilde{\omega}_I$. The real part of then frequencies 
is in general slightly larger than in GR, however, without
any possibility to discriminate between GR and pc-GR. Also
the iteration number is yet too low for going to
larger values of $-\tilde{\omega}_I$. 

However, which modes will be excited depends very much on
the dynamics in the inspiral phase. Our
suggestion is to use the numerical method as described
in \cite{rezzolla-book}, changing in the programs the
metric of GR to the one of pc-GR.

The polar modes present some practical problems,
rendering it more difficult to solve. This will be addressed
in a future publication. 

\section{Conclusions}
\label{concl}

A review was presented on the consequences of
the {\it pseudo-complex General Relativity} (pc-GR),
comparing the results with GR. In general, 
differences are only visible near massive objects 
of the size of a black hole, requiring however a large
resolution

The pc-GR is an algebraic extension of GR, where the
coordinates $x^\mu$ are redefined to
$X^\mu = x^\mu + I y^\mu$. This modifies the
Einstein equations with a dark energy energy momentum
tensor on its right hand side. This tensor depends of $u_\mu$
and ${\dot y}_\mu$. Demanding a real length element, a
differential equation for $y_\mu$ was obtained. The
$y_\mu$ are proportional to the
4-velocity, multiplied with the scalar minimal length
parameter. The $y_\mu$ was determined for the pc-Schwarzschild
case and the pc-Robertson-Walker universe. 

Observational consequences of pc-GR were determined, as the
appearance of a dark ring, followed by a bright inner ring,
in the light emission of an accretion disc.
Unfortunately
the resolution of the EHT is too low for seeing this
structure, thus, it cannot discriminate
between GR and pc-GR. 

Modification in the perihelion shift of Mercury were discussed,
with the result that pc-GR only adds corrections of 
$10^{-14}$ to the shift calculated within GR, i.e., no hope
to being seen.

Also {\it Quasi Periodic Objects} where discussed and 
interpretations within GR and pc-GR were compared, without
a definitive result.  

Neutron stars of any mass were obtained, though, for large 
masses the star rather resembles the 
one of a black hole. It is
suggested to look for light emission of infalling matter
at the poles, being excited upon impact, 
provided there is no jet emitted nearby. 
At the poles the minimal redshift is of the order of
1, increasing rapidly to infinity at the orbital plane.

The Robertson-Walker universe gives in
practice the same results as in GR for the 
present epoch and
later times. In the limit of $t \rightarrow 0$, however,
the dark energy increases significantly, implying an
important role of it at early times.

As a last example,
gravitational waves from the ring-down of a black
hole were also discussed, indicating similar results
as in GR. For large damping modes, 
the real part of the frequencies are 
larger in pc-GR than in GR.

\section*{Acknowledgments}

This work was supported by DGAPA-PAPIIT (IN100421). Very
useful discussions with Laurent R. Loinard (IRyA, UNAM)
are also acknowledged.

\end{document}